\documentclass[12pt,
tightenlines,
floats,
showpacs,
nofootinbib,
amsmath,
amssymb,
aps,
superscriptaddress,
prd]{revtex4-1}

\usepackage{amsmath,amssymb}

\begin{document}
	
\title{On the Uniqueness of the Fock Quantization of the Dirac Field in the Closed
FRW Cosmology}

\author{Jer\'onimo Cortez}
\email{jacq@ciencias.unam.mx}
\affiliation{Departamento de F\'isica, Facultad de Ciencias, Universidad Nacional Aut\'onoma de 
M\'exico, Ciudad de M\'exico  04510, M\'exico}
\author{Ilda In\'acio Rodrigues}
\email{ilda@ubi.pt}
\affiliation{Faculdade de Ci\^encias, Universidade da Beira Interior, R. Marqu\^es D'\'Avila e Bolama, 6201-001 Covilh\~a, Portugal}	
\author{Mercedes Mart\'in-Benito}
\email{mmartinb@fc.ul.pt}
\affiliation{Instituto de Astrof\'\i sica e Ci\^encias do Espa\c{c}o, Universidade de Lisboa, Faculdade de Ci\^encias, Ed. C8,
Campo Grande, PT1749-016 Lisboa, Portugal}
\affiliation{Departamento de F\'\i sica Te\'orica, Universidad Complutense de Madrid, 28040 Madrid, Spain}
\author{Jos\'e Velhinho}
\email{jvelhi@ubi.pt}
\affiliation{Faculdade de Ci\^encias, Universidade da Beira Interior, R. Marqu\^es D'\'Avila e Bolama, 6201-001 Covilh\~a, Portugal}

\begin{abstract}
The Fock quantization of free fields propagating in cosmological backgrounds is in general not unambiguously defined due to the non-stationarity of the spacetime. For the case of a scalar field in cosmological scenarios it is known that the criterion of 
unitary implementation of the dynamics serves to remove the ambiguity in the choice of Fock representation (up to unitary equivalence). Here, applying the same type of arguments and methods previously used for the scalar field case, we discuss the issue of the uniqueness of the Fock quantization  of the Dirac field in the closed FRW spacetime proposed by D'Eath and Halliwell.
\end{abstract}

\maketitle

\section{Introduction}
\label{sec:Intro}
The physics of the very early universe, and in particular relevant quantum phenomena,
can nowadays be tested, comparing {the predictions of theoretical models} against quite accurate observational data.
Besides scalar fields, it is therefore important to explore the impact of different matter sources as well (in this respect see \cite{BGM}).
Concerning fermion fields, and in particular Dirac spinors,  a consistent framework to deal with the corresponding quantum field theory in a cosmological context {was} put forward 
several years ago by D'Eath and Halliwell  \cite{H-D} (see also \cite{Dim} for a more general discussion).
 In particular, \cite{H-D} considers a concrete 
Fock quantization of the (fully inhomogeneous) Dirac field,  on a homogeneous and isotropic background spacetime, namely the closed Friedmann-Robertson-Walker
(FRW) cosmological model.   

As it is typically the case regarding the quantization of systems with an infinite number of degrees of freedom, also in the quantization of the Dirac field in a curved spacetime, one must face the issue of ambiguity,
or lack of uniqueness in the quantization procedure. Just like in the scalar field case,
the ambiguity in the quantization can be seen to lie in the choice of a so-called  {\em complex structure}
in the space of solutions (of the Dirac equation, in this case).
For the analogous problem concerning the scalar field, it was shown 
\cite{gowdy1,gowdy,prd81,spheres,AOP15} that the requirement 
of a unitary implementation of the dynamics, combined with a natural implementation of spatial symmetries,
leads to a unique  quantization (modulo unitary equivalence), thus removing the ambiguity problem.
The unitary evolution criterion replaces here the well known requirements based on invariance under 
spacetime symmetries (such as the Poincar\'e group or the symmetries of the de Sitter space) or on
stationarity, which are unavailable e.g. in typical inflationary and cosmological scenarios.
  
In the previous work \cite{PRD92}, it was shown that D'Eath and Halliwell's \cite{H-D} quantization of the Dirac field in the closed FRW cosmology satisfies  requirements similar to those just mentioned in the scalar field context, namely the complex structure chosen in  
\cite{H-D}  is invariant under the symmetries of the
field equations {(which of course include $SO(4)$, the isometry group of the spatial sections)} and admits a unitary implementation of the dynamics.
Most  importantly, it was shown that a large class of seemingly natural alternatives to the D'Eath and Halliwell's  quantization 
actually lead to the same quantization (modulo unitary equivalence) once unitary implementation of the dynamics 
is required, thus providing strong indication that a full uniqueness of quantization result is valid
for the Dirac field, in perfect correspondence with the previous results obtained for the  the scalar field case. 
{Moreover, such a uniqueness result was indeed obtained later on, as a corollary 
of {a more general analysis presented} in \cite{PRD93} (see also \cite{AOP17,PRD96}). However, the approach taken
in \cite{PRD93} was substantially different from the line of reasoning previously
adopted for the scalar field. In particular, in \cite{PRD93} two different types of ambiguities
were addressed simultaneously. In fact, besides the usual ambiguity in the choice of the quantum representation for a fixed classical field description,
time-dependent scalings of the fermion {field modes (after an appropriate mode-decomposition)} were also allowed, effectively modifying
the original D'Eath and Halliwell's parametrization of the system. 
In this respect,
a sort of time-dependent complex structures were introduced in \cite{PRD93},
 leading to a 
more effective and economic mathematical treatment.  On the other hand, 
this different perspective  might have 
contributed to a less than full appreciation of the results obtained in \cite{PRD93}.
In the current work we recover the line of reasoning followed in the scalar field case,
presenting a new and fully independent proof of part of the results of \cite{PRD93}.
In particular, we will show that, preserving D'Eath and Halliwell's field description, the quantization of the Dirac field put forward in \cite{H-D} is essentially unique,
in the sense that all Fock quantizations determined by $SO(4)$-invariant complex structures 
admitting a unitary implementation of the dynamics are unitarily equivalent.} 

Let us mention that there are similarities between the unitary evolution requirement and 
the so-called Hadamard condition, in the sense that some type
of ultraviolet regularity is imposed in both cases.
Moreover, for spacetimes with compact spatial sections, 
all (pure Fock)
Hadamard states lead to unitarily equivalent quantizations \cite{V,AH}.
We present here a different approach towards uniqueness of the quantization,
insisting on a unitary implementation of the dynamics at the quantum level. 
A brief discussion concerning the relation between these two
approaches is presented in section  \ref{sec:conclu}.

\section{The Model and its Quantization}
\label{sec:Model}

In this section we present a very brief review of the quantum treatment of the Dirac field
in {the unit three-sphere} $S^{3}$,  following
\cite{H-D} and \cite{PRD92}.

Let us consider the  closed FRW cosmological model, with metric 
\begin{align}
ds^{2}=e^{2\alpha(\eta)}(-d\eta^{2}+d\Omega_{3}^{2}),
\end{align}
where $\eta$ denotes conformal time and $d\Omega_{3}^{2}$ is the metric on {$S^{3}$}.   The massive Dirac spinor $\Psi$ is taken in the Weyl representation, i.e. is described by a  pair of  two-component spinors, $\phi^{A}$ and ${\bar{\chi}}_{A'}$,  with opposite chirality. 
The spinors $\phi_{A}$ and ${\bar{\chi}}_{A'}$ can be expanded in complete bases of spinor harmonics on $S^{3}$ provided by the eigenmodes of the Dirac operator, hereby denoted by
 $\rho_{A}^{np}$, ${\bar{\sigma}}^{np}_{A}$ and ${\bar{\rho}}^{np}_{A'}$, $\sigma_{A'}^{np}$.
Here,  $n\in\mathbb{N}$, the degeneracy of each eigenspace is 
\begin{align}
g_n=(n+1)(n+2)=\omega_n^2-\frac{1}{4}, 
\end{align} 
with $\omega_{n}=n+3/2$, and 
the label $p=1,...,g_n$ accounts for the degeneracy of the eigenspaces.

Explicitly, the two-component spinors  (and their Hermitian conjugates) can be expanded in modes as   follows:
\begin{align}\label{harm1}
\phi_A(x)=&\frac{e^{-3\alpha(\eta)/2}}{2\pi}\sum_{npq}\breve{\alpha}_{n}^{pq}[m_{np}(\eta)\rho_{A}^{nq}(\textbf{x})+{\bar{r}}_{np}(\eta){\bar\sigma}_{A}^{nq}(\textbf{x})],
\\ \label{harm4}
{\bar\chi}_{A'}(x)=&\frac{e^{-3\alpha(\eta)/2}}{2\pi}\sum_{npq}\breve{\beta}_{n}^{pq}[{\bar{s}}_{np}(\eta){\bar\rho}_{A'}^{nq}(\textbf{x})+t_{np}(\eta)\sigma_{A'}^{nq}(\textbf{x})],
\end{align}
with
\begin{align}
\sum_{npq}:=\sum_{n=0}^{\infty}\sum_{p=1}^{g_n}\sum_{q=1}^{g_n}.\nonumber
\end{align}
The Grassmann variables $m_{np}$, $r_{np}$, $s_{np}$, $t_{np}$ therefore describe the fermionic degrees of freedom, after mode decomposition. {In the above expressions, the coefficients $\breve{\alpha}_{n}^{pq}$ and $\breve{\beta}_{n}^{pq}$ are the matrix elements of two constant matrices $\breve{\alpha}_{n}$ and $\breve{\beta}_{n}$, each of dimension $g_n$ and block-diagonal form, with blocks given respectively by
\begin{align}
 \begin{pmatrix}
  1 & 1\\
  1& -1
 \end{pmatrix}\quad\text{and}\quad  
 \begin{pmatrix}
  1 & -1\\
  -1& -1
 \end{pmatrix}.
\end{align}
They are included by convenience, to avoid dynamical couplings between mode components with different values of $p$.
}

The mode components of the  Dirac equation,  deduced from the Einstein-Dirac action,  can be summarized in the following set of differential  equations (and corresponding complex conjugates) \cite{H-D}:
\begin{align}\label{1order}
x_{np}'=i\omega_{n}x_{np}-ime^{\alpha}\bar{y}_{np}, \qquad y_{np}'=i\omega_{n}y_{np}+ime^{\alpha}\bar{x}_{np}.
\end{align}
In the above equations,  $m$ stands for the mass of the fermion field, and the prime denotes  derivative with respect to conformal time.  We are adopting here the notation $(x_{np},y_{np})$ to denote  any of the sets of  pairs  $(m_{np},s_{np})$ or $(t_{np},r_{np})$.

As mentioned in the introduction, the ambiguity in the Fock quantization of the Dirac field resides in the choice of a complex structure in the space of solutions of the Dirac equation.
This in turn amounts to  a choice of a set of classical annihilitation and creation-like variables, to be quantized as  annihilitation and creation operators (see \cite{PRD92} for details).

The preferred choice of annihilation and creation-like variables considered in  {\cite{H-D} diagonalizes the Hamiltonian and} is defined as follows.
The annihilation-like variables of the particles and antiparticles are  chosen to be
{\begin{align}\label{refcs}
a^{(x,y)}_{np}=\sqrt{\frac{1}{2}-\frac{1}{2\xi_n}} x_{np}+\sqrt{\frac{1}{2}+\frac{1}{2\xi_n}}\bar{y}_{np}, \qquad b^{(x,y)}_{np}=\sqrt{\frac{1}{2}+\frac{1}{2\xi_n}}\bar{x}_{np}-\sqrt{\frac{1}{2}-\frac{1}{2\xi_n}}y_{np},
\end{align}
with 
\begin{equation}
\xi_n=\sqrt{1+\left(\frac{me^\alpha}{\omega_n}\right)^2}.
\end{equation}}

The creation-like variables $a_{np}^{(x,y)\dagger}:=\bar{a}_{np}^{(x,y)}$ and $b_{np}^{(x,y)\dagger}:=\bar{b}_{np}^{(x,y)}$ are their complex conjugates. When a complete Hamiltonian analysis of the Einstein-Dirac action is performed, one finds that  these variables indeed satisfy the Dirac brackets characteristic of annihilation and creation-like variables for particles and antiparticles, namely
\begin{align}\label{CACR}
\{a^{(x,y)}_{np},a^{(x,y)\dagger}_{np}\}=
\{b^{(x,y)}_{np},b^{(x,y)\dagger}_{np}\}=-i,\qquad \{a^{(x,y)}_{np},b^{(x,y)}_{np}\}=0.
\end{align}

Let us call {\em reference complex structure} and {\em reference quantization} the ones determined by the choice of variables  introduced above.

The main feature of this quantization is that it allows a unitary implementation of the
classical field dynamics in the quantum theory. Let us see exactly what this means. 
Since both the equations of motion and the relations \eqref{refcs} are linear, it is clear that the variables
defined by  relations \eqref{refcs} evolve linearly in time. In particular,  the evolution from an
arbitrary  initial time $\eta_0$ 
to any other time $\eta$ is given by a Bogoliubov transformation of the form
\begin{align}\label{bog}
\begin{pmatrix} a_{np}^{(x,y)} \\ b^{(x,y)\dagger}_{np} \end{pmatrix}_{\!\!\eta}=B_{n}(\eta,\eta_{0})\begin{pmatrix} a_{np}^{(x,y)} \\ b^{(x,y)\dagger}_{np} \end{pmatrix}_{\!\!\eta_{0}},
\end{align}
with
\begin{align}\label{bog1}
B_{n}=\begin{pmatrix} \alpha_{n}^{f} & \beta_{n}^{f} \\ \beta_{n}^{g} & \alpha_{n}^{g} \end{pmatrix},
\end{align}
where $\alpha_{n}^{f}$ and $\beta_{n}^{f}$ are coefficients (dependent  on $\eta$ and $\eta_0$) whose full 
expression can be found in  \cite{PRD92}.

It turns out that  a classical transformation of the type (\ref{bog}) can be unitarily implemented
(in the Fock representation defined by the complex structure corresponding to the choice of variables 
(\ref{refcs})) if and only if \cite{Shale}  the sum
\begin{align}\label{z1}
\sum_{n}g_n(|\beta^{f}_{n}|^{2}+|\beta^{g}_{n}|^{2})
\end{align}
is finite. Of course, since in the present case one is interested in the unitary
implementation of the dynamics, one must require that the above sum be finite for all values of
time   $\eta$ and $\eta_0$, which was proven to be the case \cite{PRD92}.

\section{Alternative Complex Structures and Unitarity Condition}

The question we now want to answer is the following: is the quantization described in the previous section unique? In what sense? We have learned from the analysis of similar cases involving scalar fields, that the requirement of unitary implementation of the dynamics is a successful criteria
in the selection of quantum representations. In fact, together with the requirement of invariance
of the complex structure under remaining spatial symmetries, the unitary dynamics condition
selects a unique quantization (modulo unitary equivalence) for the scalar field propagating 
in a nonstationary homogeneous and isotropic 
spacetime background \cite{spheres}, typical of cosmological scenarios.  
Note that the invariance of the complex structure guarantees a natural unitary implementation of the symmetries in question, and it is therefore 
 reasonable to restrict attention to those invariant complex structures.
Of course, {our considerations} should be extended to noninvariant complex structures
whenever the physical conditions require it or if physically relevant
noninvariant complex structures are known to exist. 
That is not the case in the present situation, to the best of our knowledge. On the other hand,
the unitary implementation  of the dynamics cannot be made via complex structures 
that remain invariant under evolution, for the simple reason that such complex structures
do not exist, due to the lack of stationarity.

As shown in detail in \cite{PRD93}, the general form of a complex structure that remains invariant  under
the action of the isometry group of $S^3$ on
the space of two-component spinors
  is very simple to characterize. Such an invariant complex structure is associated with a different choice of annihilation and creation-like variables {$\{\tilde{a}_{np}^{(x,y)},\tilde{b}_{np}^{(x,y)},\tilde{a}^{(x,y)\dagger}_{np},\tilde{b}^{(x,y)\dagger}_{np}\}$}, related to our reference variables 
(\ref{refcs}) by means of a {transformation (and its complex conjugate)} of the type
\begin{align}\label{newCS}
\begin{pmatrix} \tilde{a}_{np}^{(x,y)} \\ \tilde{b}^{(x,y)\dagger}_{np}\end{pmatrix}=K_{n}\begin{pmatrix} a_{np}^{(x,y)} \\ b^{(x,y)\dagger}_{np} \end{pmatrix},
\end{align}
with
\begin{align}
\label{n32}
K_{n}=\begin{pmatrix} \kappa_{n}^{f} & \lambda_{n}^{f} \\ \lambda_{n}^{g} & 
\kappa_{n}^{g} \end{pmatrix},
\end{align}
where  the time-independent matrices $K_{n}$, $n\in\mathbb{N}$, are arbitrary  
$2\times 2$ {unitary} matrices, i.e. they  satisfy the following relations:
\begin{align}\label{orto}
|\kappa_{n}^{f}|^{2}+|\lambda_{n}^{f}|^{2}=1, \qquad |\kappa_{n}^{g} |^{2}+|\lambda_{n}^{g}|^{2}=1, \qquad \kappa_{n}^{f}\bar{\lambda}^{g}_{n}+\lambda_{n}^{f}\bar{\kappa}^{g}_{n}=0,
\end{align}
as follows from \eqref{CACR}.


Among the large class of quantum representations of the Dirac field determined by the 
invariant complex structures as above, there is certainly an infinite number of representations
that are not unitarily equivalent to our reference quantization. In fact, the condition 
for unitary equivalence reads \cite{Shale,Der}
\begin{align}\label{newucond}
\sum_{n}g_{n}(|\lambda_{n}^{f}|^{2}+|\lambda_{n}^{g}|^2)<\infty, 
\end{align}
which is obviously not satisfied by an arbitrary sequence of matrices $K_n$.

{At this point, it should be mentioned that there is an important difference 
-- first noted in  \cite{PRD92} -- with respect to the previously studied
scalar field case, that we will now  address. Note that orthogonality conditions 
\eqref{orto} (contrary to corresponding symplectic conditions in the scalar field case) 
can be fulfilled with $\lambda_{n}^{f}=\lambda_{n}^{g}=1$,  $\kappa_{n}^{f}=\kappa_{n}^{g}=0$.
It is clear that a transformation of this type simply interchanges  particles
with antiparticles, and therefore has no physical meaning, since that distinction is conventional to begin with. However, if one allows
such a transformation for an infinite number of modes, one ends up with two formally
inequivalent quantizations, since the equivalence condition \eqref{newucond}
is clearly violated. In order to eliminate the artificially inequivalent complex structures associated with these type of transformations, we will adhere to a fixed convention as to what
is called particle and antiparticle, which we will not allow to change (except possibly for a finite number of modes).  A concrete implementation of this notion is to require e.g. that 
the sequences $\{\kappa_{n}^{f}\}$ and $\{\kappa_{n}^{g}\}$ in \eqref{n32} have no subsequences that tend to zero, which we will assume from now on. There still remains, of course, a large class of physically meaningful complex structures.}

We are going to show, precisely, that once unitary implementation of the dynamics  is imposed
as a requirement,
only those complex structures that satisfy (\ref{newucond}) survive. As a preparation, let 
us state a preliminary consequence of the existence of unitary dynamics.

One can easily show (see \cite{gowdy1}) that  a complex structure associated with a choice
of variables (\ref{newCS}), determined by matrices $K_n$, allows a unitary implementation 
of the dynamics if and only if the time-dependent transformation determined by the 
sequence of matrices $K_nB_n(\eta,\eta_0)K_n^{-1}$ is unitarily implementable with respect to
our reference quantization. A straightforward computation shows that, as a consequence,
the following two sequences must be square summable, $\forall \eta$:
\begin{align}
\sqrt{g_{n}}\Bigl((\kappa_{n}^{f})^{2}\beta_n^f-(\lambda_{n}^{f})^2\beta_n^g+\kappa_{n}^{f}\lambda_{n}^{f}(\alpha_n^g-\alpha_n^f)\Bigr), 
\end{align}
and
\begin{align}
\sqrt{g_{n}}\Bigl((\kappa_{n}^{g})^{2}\beta_n^g-(\lambda_{n}^{g})^2\beta_n^f+\kappa_{n}^{g}\lambda_{n}^{g}(\alpha_n^f-\alpha_n^g)\Bigr).
\end{align}
Since we know that {\em i}\/) the sequences $\sqrt{g_{n}}\beta_n^f$ and $\sqrt{g_{n}}\beta_n^g$
are square summable, because the sum (\ref{z1})  was shown to be finite \cite{PRD92}, and
{\em ii}\/) the sequences $\kappa_{n}$ and $\lambda_{n}$ are bounded, it follows (from linearity
of the set of square summable sequences) that necessarily 
\begin{align}\label{z2}
\sum_n g_{n}|\kappa_{n}^{f}\lambda_{n}^{f}(\alpha_n^g-\alpha_n^f)|^2<\infty,
\end{align}
and
\begin{align}\label{z3}
\sum_n g_{n}|\kappa_{n}^{g}\lambda_{n}^{g}(\alpha_n^f-\alpha_n^g)|^2<\infty.
\end{align}
These conditions are therefore consequences of the requirement of {unitarity of the} dynamics.
We will take these as the starting point for our next section, were we will prove that
conditions \eqref{z2}-\eqref{z3} in fact imply that \eqref{newucond} is satisfied. 

\section{Uniqueness of the Quantization}
A detailed asymptotic analysis of the evolution matrices {$B_{n}$} \eqref{bog1} was performed in  \cite{PRD92}.
Following the same procedure that allowed to obtain an expression 
for $\beta_n^h$ in the limit of large $n$ (equation (3.18) in  \cite{PRD92}),
we obtain the following expression for $\alpha_n^g-\alpha_n^f$, $\forall \eta$,  in the limit of large $n$:
{\begin{align}
(\alpha_{n}^{g}-\alpha_{n}^{f})&(\eta,\eta_0)=2i
\bigg\{\big[f_{2}^{n}(\eta)f_{2}^n(\eta_0)-f_{1}^{n}(\eta)f_{1}^{n}(\eta_0)\big]e^{-\Im(\Lambda_{n}^{1}(\eta))}
\sin\Big(\omega_n\Delta\eta+\int_{\eta_0}^{\eta}\Re\big(\Lambda_{n}^{1}(\bar\eta)\big)d\bar\eta\Big)
\nonumber \\ 
+&\big[f_{1}^{n}(\eta)f_{2}^{n}(\eta_0)+f_{2}^{n}(\eta)f_{1}^{n}(\eta_0)\big]|\Gamma_n|e^{\Delta\alpha}\Big[
e^{-\Im(\Lambda_{n}^{1}(\eta))}
\sin\Big(\omega_n\Delta\eta+\int_{\eta_0}^{\eta}\Re\big(\Lambda_{n}^{1}(\bar\eta)\big)d\bar\eta
+\varphi_n\Big)\nonumber \\ 
-&e^{\Im(\Lambda_{n}^{2}(\eta))}
\sin\Big(\omega_n\Delta\eta+\int_{\eta_0}^{\eta}\Re\big(\Lambda_{n}^{2}(\bar\eta)\big)d\bar\eta
-\varphi_n\Big)\Big]\bigg\}.
\end{align}}
In the above expression we have $\Delta\eta=\eta-\eta_{0}$, 
$\Delta\alpha=\alpha(\eta)-\alpha(\eta_{0})$, 
\begin{align}
f_1^n&=\sqrt{\frac{1}{2}-\frac{1}{2\xi_n}}=\frac{me^{\alpha}}{2\omega_n}+\mathcal{O}(\omega_{n}^{-2}),\\
 f_2^n&=\sqrt{1-|f_1^n|^2}=1+\mathcal{O}(\omega_{n}^{-2}),\\
\Gamma_n&=\frac{me^{\alpha(\eta_0)}}{2\omega_n+i\alpha'(\eta_0)}=\frac{me^{\alpha(\eta_0)}}{2\omega_n}+\mathcal{O}(\omega_{n}^{-2}),
\end{align}
$\varphi_n$ is the phase of $\Gamma_n$, and 
$\Lambda^{1}_{n}$  and $\Lambda^{2}_{n}$ are time-dependent functions, defined in the appendix of 
 \cite{PRD92}, which have the property of being of order $\mathcal{O}(\omega_{n}^{-1})$ in the limit of large $n$. 

Taking into account the asymptotic limits of the coefficients $f_1^n$,  $\Gamma_n$, 
$\Lambda^{1}_{n}$  and $\Lambda^{2}_{n}$, it follows that all the terms 
in $\sqrt{g_{n}}\kappa_{n}^{f}\lambda_{n}^{f}(\alpha_n^g-\alpha_n^f)$ are square
summable, with the possible exception of the term 
\begin{equation}\label{z4}
2i\sqrt{g_{n}}\kappa_{n}^{f}\lambda_{n}^{f}f_{2}^{n}(\eta)f_{2}^{n}(\eta_0)e^{-\Im(\Lambda_{n}^{1})}
\sin\Big(\omega_n\Delta\eta+\int_{\eta_0}^{\eta}\Re\big(\Lambda_{n}^{1}(\bar\eta)\big)d\bar\eta\Big).
\end{equation}

However, since we are assuming that the sequence   $\sqrt{g_{n}}\kappa_{n}^{f}\lambda_{n}^{f}(\alpha_n^g-\alpha_n^f)$ itself is square
summable, it follows  that the sequence \eqref{z4} is square summable as well.
Moreover, since $f_{2}^{n}(\eta)f_{2}^{n}(\eta_0)e^{-\Im(\Lambda_{n}^{1})}$ actually tends to one in the 
large $n$ limit, we conclude that
\begin{equation}\label{z5}
\sum_n g_n|\kappa_{n}^{f}\lambda_{n}^{f}|^2\sin^2\Big(\omega_n\Delta\eta+\int_{\eta_0}^{\eta}\Re\big(\Lambda_{n}^{1}(\bar\eta)\big)d\bar\eta\Big)<\infty.
\end{equation}
Obviously, the same reasoning leads to the conclusion that
\begin{equation}\label{z6}
\sum_n g_n|\kappa_{n}^{g}\lambda_{n}^{g}|^2\sin^2\Big(\omega_n\Delta\eta+\int_{\eta_0}^{\eta}\Re\big(\Lambda_{n}^{1}(\bar\eta)\big)d\bar\eta\Big)<\infty.
\end{equation}
These two conditions are therefore   consequences of \eqref{z2}-\eqref{z3}, which are in turn consequence
of our requirement of unitary implementation of the dynamics. 
From this point on, the proof that \eqref{z5}-\eqref{z6} implies 
\eqref{newucond} can be done along the lines of the argument
presented in  \cite{prd81}.
Let us then introduce  the shifted time
$T:=\eta-\eta_0$, and rewrite the elements of the sequence appearing in 
\eqref{z5} in the form
\begin{equation}
\label{44} \sqrt{g_n} \,{\kappa^f_n}{\lambda^f_n}
\,\sin{\left[ \omega_nT+\int_{0}^T
d\bar{\eta}\,\Re\big(\Lambda_{n}^{1}(\bar\eta+\eta_0)\big)\right]}.
\end{equation}
The function
\begin{equation}
\label{45} z(T):=\lim_{M\to
\infty}\sum_{n=1}^{M}g_n{|\kappa^f_n|^{2}}{|\lambda^f_n|^{2}}
\sin^{2}{\left[ \omega_nT+\int_{0}^T
d\bar{\eta}\,\Re\big(\Lambda_{n}^{1}(\bar\eta+\eta_0)\big)\right]}
\end{equation}
therefore exists for all $T$ 
{in the domain 
of $\eta-\eta_0$. In particular, $z(T)$ is well defined
on some closed subinterval 
$\mathbb{\bar{I}}_L=[a,a+L]$,
 where $L$ is
some finite positive number.}

Luzin's theorem \cite{luzin} guarantees that, for
every $\delta>0$, there exist: i) a measurable set
$E_{\delta} \subset \mathbb{\bar{I}}_L$ such that its
complement $\bar{E}_{\delta}$ with respect to
$\mathbb{\bar{I}}_L$ satisfies
{$\int_{\bar{E}_{\delta}}d T<\delta$}, and ii) a
function $F_{\delta}(T)$ which is continuous on
$\mathbb{\bar{I}}_L$ and coincides with $z(T)$ in
$E_{\delta}$. In particular, defining
$I_{\delta}:=\int_{E_{\delta}}F_{\delta}(T)d T$,
 we obtain from Luzin's
theorem:
\begin{equation}
\label{46}
\sum_{n=1}^{M}g_n{|\kappa^f_n|^{2}}{|\lambda^f_n|^{2}}
\int_{E_{\delta}}\sin^{2}{\left[ \omega_nT+\int_{0}^T
d\bar{\eta}\,\Re\big(\Lambda_{n}^{1}(\bar\eta+\eta_0)\big)\right]}dT\leq \int_{E_{\delta}}z(T)
d T=I_{\delta}, \quad
\forall M \in \mathbb{N}^+ .
\end{equation}
This inequality provides us
with a bound on $\sum g_n
{|\kappa^f_n|^{2}}{|\lambda^f_n|^{2}}$. To show it, we note
that, $\forall n$:
\begin{align}
\label{47} \nonumber \int_{E_{\delta}}\sin^{2}{\left[
\omega_nT+\int_{0}^T
d\bar{\eta}\,\Re\big(\Lambda_{n}^{1}(\bar\eta+\eta_0)\big)\right]}d
T \\
\nonumber =\int_{\mathbb{\bar{I}}_L} \sin^{2}{\left[
\omega_nT+\int_{0}^T
d\bar{\eta}\,\Re\big(\Lambda_{n}^{1}(\bar\eta+\eta_0)\big)\right]} d
T-&\int_{\bar{E}_{\delta}}\sin^{2}{\left[
\omega_nT+\int_{0}^T
d\bar{\eta}\,\Re\big(\Lambda_{n}^{1}(\bar\eta+\eta_0)\big)\right]} d T
\\ \geq\int_{\mathbb{\bar{I}}_L}\sin^{2}{\left[
\omega_nT+\int_{0}^T
d\bar{\eta}\,\Re\big(\Lambda_{n}^{1}(\bar\eta+\eta_0)\big)\right]} d T
-&\delta.
\end{align}
 Applying now an integration by parts and  a bound just like  in the appendix
of \cite{prd81}, we find the following for the integral over
${\mathbb{\bar{I}}_L}$:
\begin{equation}
\int_{\mathbb{\bar{I}}_L}  \sin^2\left[
\omega_n T+\int_{0}^T d\bar{\eta}\,\Re\big(\Lambda_{n}^{1}(\bar\eta+\eta_0)\big)\right]d T \geq
\frac{L}{2} - \frac{1}{2 \omega_{n_0} (1-D)} -\frac{
\int_{\mathbb{\bar{I}}_L}  |C'(T+\eta_0)|d T}{4
\omega_{n_0}^3 (1-D)^2}:= \Lambda_{n_0}.
\end{equation}
Here, $C$ is the coefficient of the term in $\omega_n^{-1}$ of the Laurent series expansion 
of  $\Re(\Lambda_{n}^{1})$.
This expression is valid for  $n\geq
n_0$, where $n_0$ is any fixed (positive) integer such
that $\omega_{n_0}^2$ is larger than the maximum of
the function $|C(T+\eta_0)|/(2D)$ in the interval
${\mathbb{\bar{I}}_L}$, and $D<1$ is any fixed
constant. We are
 assuming also that $n_0$ is such that $\Lambda_{n_0}>0$
(which can certainly be achieved with an appropriate
choice, since $\Lambda_{n_0}$ tends to $L/2$ when
$n_0$ tends to infinity).

Let us introduce the above result in the last
inequality of  (\ref{47}), to obtain:
\begin{equation}
\int_{E_{\delta}}\sin^{2}{\left[ \omega_nT+\int_{0}^T
d\bar{\eta}\,\Re\big(\Lambda_{n}^{1}(\bar\eta+\eta_0)\big)\right]} d T
\geq \Lambda_{n_0}-\delta.
\end{equation}
We choose now $\delta$ such that $\Lambda_{n_0}>\delta$,
 which is certainly possible. Then, it follows
from  (\ref{46}) that, for all $M\geq n_0$,
\begin{equation}
\sum_{n=n_0}^M g_n
|\kappa^f_n|^2|\lambda^f_n|^2 \leq \frac{
I_{\delta}}{\Lambda_{n_0}-\delta}.
\end{equation}
Since $n_0$ is fixed and the above bound is true for
arbitrarily large $M$, it follows that the sequence 
 $\sqrt{g_n}\kappa^f_n\lambda^f_n$
is square summable.

{At this point, let us recall from the discussion in the previous section that,
without loss of physical generality, the sequence  $\{\kappa^f_n\}$ can be assumed to be bounded from below. Thus, it follows moreover that the sequence 
 $\sqrt{g_n}\lambda^f_n$
is square-summable. Finally, since obviously the same reasoning applies to (\ref{z6}),
one is lead to the conclusion that condition (\ref{newucond}) is satisfied.
It is therefore proven that the quantization of the Dirac field put forward
in Ref. \cite{H-D} is indeed unique {up to unitary equivalence}, under the requirements of invariance of the complex
structure and unitary implementation of the dynamics.} 

\section{Conclusions and Discussion}
\label{sec:conclu}
We have analyzed the issue of the uniqueness of the Fock quantization 
of the Dirac field in the closed FRW spacetime proposed by D'Eath and Halliwell  \cite{H-D}.
We have worked with a fixed parametrization of the system, leaving aside the extra
freedom that resides in the possibility of applying (to the variables to be quantized)
further time-dependent transformations. It was therefore possible to address the uniqueness
issue  in exactly the same way as previously done for the scalar field case.
In this sense, the results here obtained are more restricted than those of \cite{PRD93},
but were proven in a fully independent way, using arguments that are more familiar to
the quantum cosmology literature. It was fully confirmed that, once a convention for
particles-antiparticles is agreed upon, the D'Eath and Halliwell's Fock quantization is indeed
unique, subjected to the requirements of  invariance of the complex structure
under the group of spatial isometries and unitary implementation
of the dynamics in Fock space.

A question that deserves discussion is the following.
It has been argued by several authors that so-called Hadamard states and corresponding
quantum representations are physically privileged for quantum field theory
in curved spacetime since, in particular, they allow a regularization of
the  stress-energy tensor and a well defined (perturbative) construction of 
interacting theories. Moreover, it is known -- as a consequence of more general results
in \cite{AH} and following previous results for the scalar field -- 
that  for the Dirac field
in a globally hyperbolic spacetime with compact spatial sections,
all pure Fock Hadamard states lead to unitarily equivalent quantizations.
In the special case of cosmological spacetimes, such as the case considered
in the present paper, another privileged family of states is known,
that of adiabatic states, and again all adiabatic states  (of sufficiently high 
order)  give rise to unitarily equivalent representations (in the spatially compact case) 
\cite{H}.
Morevover, adiabatic states  (of sufficiently high order) and Hadamard states belong to the same
unitary equivalence class  (in the spatially compact case).
Thus, criteria leading to unitarily equivalent quantizations (in the spatially compact case)
do exist, and in that strict sense the results of the current paper bring essentially no 
novelty, except that of an alternative approach.

A similar question emerged in the previous studies of analogous -- e.g. using the criteria
of unitary implementation of the dynamics -- uniqueness of
quantization results concerning the scalar field \cite{gowdy1,gowdy,prd81,spheres,AOP15}.
It is again well known  that  for the scalar field
in a globally hyperbolic spacetime (with compact spatial sections)
all pure Fock Hadamard states give rise to unitarily equivalent quantizations
\cite{V}. 
For the case of the scalar field in the closed FRW spacetime, the relation between 
the quantum representation selected
by the unitary evolution requirement and the one provided by Hadamard states was discussed in some
detail in \cite{spheres}, with the following conclusions. 
Once the  time
dependent scaling $\phi \to \varphi =a\phi$ is taken into account, where $a$ is the FRW scale factor,
the two criteria lead to equivalent quantizations. In more precise terms, the unique
(unitary equivalence class of) quantum representation determined by the requirement of
unitary implementation of the classical dynamics of the field $\varphi$,  induces
a Fock quantization of the  original field $\phi$ which is unitarily equivalent to the 
ones associated to both adiabatic and Hadamard states.
(See \cite{JS} for a general definition of adiabatic states
and  \cite{LR,J} for the special FRW setting.
In the particular case of compact spatial sections it can be shown that
all  adiabatic and Hadamard states give rise to unitarily equivalent representations 
\cite{J}.) 

If one believes, as we do,
that preserving unitarity of the dynamics as much as possible is a desirable
aspect in quantum physics, the fact that this perspective
actually leads to a quantum theory that is equivalent to the one  
associated with the celebrated Hadamard states, appears
by itself as an interesting and reassuring result.

Based on the previous experience with the scalar field, we  likewise expect that,
in the current context of the Dirac field in the closed FRW spacetime, both Hadarmard
states and the requirement of unitary implementation of the dynamics would lead to essentially equivalent
quantizations. However, the detailed analysis of the relation between the two 
different approaches in the Dirac field case seems rather involved, in comparison 
with the previous study concerning the scalar field [7], and it falls outside
the scope of the present work.

\acknowledgments
The authors are immensely grateful to B. Elizaga Navascu\'es and G. A. Mena Marug\'an for their help, insight and constant support.
This work was partially supported by the research grants 
MICINN/MINECO Project No. FIS2014-54800-C2-2-P from Spain, 
DGAPA-UNAM IN113115 and CONACyT 237351 from Mexico. J. V. acknowledges the COST Action CA16104 GWverse, supported by COST (European Cooperation in Science and Technology). In addition, M. M-B. acknowledges financial support from the Portuguese Foundation for Science and
Technology (FCT, Grant No. IF/00431/2015).


\end{document}